\documentclass[11pt]{article}

\usepackage{graphicx}
\begin{document}
\oddsidemargin .03in
\evensidemargin 0 true pt
\topmargin -.4in


\def\ra{{\rightarrow}}
\def\a{{\alpha}}
\def\b{{\beta}}
\def\l{{\lambda}}
\def\eps{{\epsilon}}
\def\T{{\Theta}}
\def\t{{\theta}}
\def\co{{\cal O}}
\def\car{{\cal R}}
\def\caf{{\cal F}}
\def\cs{{\Theta_S}}
\def\pr{{\partial}}
\def\tri{{\triangle}}
\def\na{{\nabla }}
\def\S{{\Sigma}}
\def\s{{\sigma}}
\def\sp{\vspace{.15in}}
\def\hs{\hspace{.25in}}

\newcommand{\be}{\begin{equation}} \newcommand{\ee}{\end{equation}}
\newcommand{\bea}{\begin{eqnarray}}\newcommand{\eea}
{\end{eqnarray}}


\begin{titlepage}
\topmargin= -.2in
\textheight 8.8in

\begin{center}
\baselineskip= 20 truept

\vspace{.3in}
\centerline{\Large\bf Quintessence and effective AdS brane geometries}

\vspace{.6in}
\noindent
{{\bf K. Priyabrat Pandey}, {\bf Abhishek K. Singh}, {\bf Sunita Singh} {\bf and} {{\bf Supriya Kar}\footnote{skkar@physics.du.ac.in }}}

\vspace{.2in}

\noindent


\noindent
{{\Large Department of Physics \& Astrophysics}\\
{\Large University of Delhi, New Delhi 110 007, India}}

\vspace{.2in}

{\today}
\thispagestyle{empty}

\vspace{.6in}
\begin{abstract}
A geometric torsion dynamics leading to an effective curvature in a second order formalism on a $D_4$-brane is revisited with a renewed interest.
We obtain two effective $AdS_4$ brane geometries on a vacuum created pair of $(D{\bar D})_3$-brane. One of them is shown to describe an AdS Schwarzschild spinning black hole and the other is shown to describe a spinning black hole bound state. It is argued that a D-instanton in a vacuum created anti $D_3$-brane within a pair may describe a quintessence. It may seen to incorporate a varying vacuum energy density in a brane universe.
We consider the effective curvature scalar on $S^1\times S^1$ to analyze torsion-less geometries on a vacuum created pair of $(D{\bar D})_2$-brane. The emergent $AdS_3$ brane is shown to describe a Schwarzschild and a Reissner-Nordstrom (RN) geometries in presence of extra dimension(s).

\baselineskip=16 truept

\vspace{.12in}

\vspace{1in}
\noindent

\noindent
\end{abstract}
\end{center}

\vspace{.2in}

\baselineskip= 16 truept

\vspace{1in}



\end{titlepage}

\baselineskip= 18 truept

\section{Introduction}
The pair production mechanism, underlying a particle and an anti-particle, at a black hole horizon has been established as a powerful tool to address some of the quantum gravity effects in the folklore of theoretical physics \cite{hawking}. The field theoretic idea was generically applied in presence of a non-trivial background metric leading to a black hole. In particular a photon in a $U(1)$ gauge theory is known to produce an electron-positron pair 
at a black hole event horizon which in turn is believed to explain the Hawking radiation phenomenon. In addition to a number of established semi-classical black holes in ($3+1$) and higher dimensions, a ($2+1$) dimensional spinning Anti de Sitter (AdS) black hole in Einstein gravity has 
been addressed in the past \cite{BTZ,BHTZ}. The results convincingly established a ($2+1$) dimensional Einstein gravity in presence of a constant vacuum energy density and a spin angular momentum. In the context a background NS two form is known to introduce a torsion in the low energy string effective action \cite{candelas-HS,freed}. Interestingly a stringy black hole in $4D$ has been obtained in a low energy string effective action \cite{garfinkle-HS}.

\sp
\noindent
In the context Dirichlet (D) $p$-branes are believed to possess a potential tool to explore a new vacuum in a superstring theory. A $D_p$-brane is
charged under their Ramond-Ramond (RR) forms and they are non-perturbative objects propagating in a perturbative string theory \cite{polchinski}.
A strongly coupled gauge theory on a $D_3$-brane has been conjectured as a dual to a weakly coupled $AdS_5$ gravity \cite{maldacena,witten}. In particular a $D_3$-brane is believed to describe the boundary of an $AdS_5$ in string theory. The idea AdS/CFT, underlying an AdS gravity in bulk and a gauge theory at its boundary, has been extended to investigate a correspondence between a de Sitter (dS) gravity and gauge theory
\cite{strominger-ds, bousso-maloney-strominger}. Various interesting notions such as lower dimensional branes within a higher dimensional $D$-brane \cite{douglas-branes}, a discrete torsion \cite{douglas-torsion}, non-linear gauge theory on a $D$-brane \cite{seiberg-witten}, near horizon geometries on a $D$-brane \cite{gibbons}-\cite{nicolini2} and string cosmology underlying a pair of $D$-brane/anti-brane universe \cite{burgess-majumdar,majumdar-davis,keshav-radu} have intensely been discussed in literatures.

\sp
\noindent
In the recent past, authors have argued for an effective space-time created on a pair of $(D{\bar D})_3$-brane by a dynamical two form in a $U(1)$ gauge theory on a $D_4$-brane \cite{spsk-JHEP}-\cite{psskk}. Primarily a non-linear gauge dynamics on a $D_4$-brane is explored to construct a geometric torsion underlying an effective curvature scalar in a second order formalism. The Poincare duality has played a significant role to identify a gauge theoretic torsion $H_3$ underlying the $U(1)$ gauge symmetry on a $D_4$-brane \cite{kpss}. In particular we have considered the two form gauge dynamics in presence of a non-trivial background metric on a $D_4$-brane. Importantly the background is identified with an open string effective metric which in turn is sourced by a global mode of NS two form in an open string bulk \cite{seiberg-witten}.
Presumably the setup, in presence of background open string metric, hints at an anti $D_4$-brane in the formalism. A pair of brane and anti-brane breaks supersymmetry and may describe a non-BPS configuration. A propagating geometric torsion in the formalism is shown to incorporate the global modes of a NS two form in addition to a local two form on a $D_4$-brane. The two form $U(1)$ gauge invariance of the effective curvature scalar ${\tilde{\cal K}}$ has been shown to ensure a metric fluctuation which in turn is believed to describe an effective space-time on a $D$-brane. Interestingly a cosmological pair production of $D{\bar D}$-brane has been discussed in a different context \cite{majumdar-davis}.

\sp
\noindent
In the paper we focus on some of the effective AdS brane geometries on a vacuum created pair of $(D{\bar D})_3$-brane and on a pair of $(D{\bar D})_2$-brane. In particular we obtain two distinct $AdS_4$ brane universe described by a quantum Schwarzschild black hole and a black hole bound state.
Interestingly an AdS brane in a different context has been discussed in ref.\cite{bachas}. We show that a torsion incorporates a non-linear axionic charge into a brane universe which in turn is believed to source a D-instanton. An instanton on a vacuum created anti $D_3$-brane, within a pair, may describe a quintessence. It may seen to incorporate a varying vacuum energy density in a brane universe. Interestingly quintessence in string theory has been investigated in the past \cite{susskind-HK}. For recent articles, see refs. \cite{chen-jing-quintessence}-\cite{huan-jun-quintessence}.

\sp
\noindent
Furthermore we consider the effective curvature scalar ${\tilde{\cal K}}$ on $S^1\times S^1$ and investigate the space-time on a vacuum created pair of $(D{\bar D})_2$-brane. We obtain an $AdS_3$ brane with a Schwarzschild black hole and a RN black hole on a $D_2$-brane universe. The significance of a dynamical axion on a $D_2$-brane or on an anti-brane is analyzed to hint at the source of dark energy underlying an accelerating brane universe.

\sp
\noindent
The plan of the article is the following. We begin with a brief introduction to an effective torsion curvature formalism in section 1. We primarily address some of the effective AdS geometries obtained on a vacuum created pair of $(D{\bar D})_3$-brane in section 2 and subsequently on a pair $(D{\bar D})_2$-brane in section 3. The effective geometries are analyzed to obtain (i) an $AdS_4$ Schwarzschild black hole in section 2.1, an $AdS_4$ quantum black hole in section 2.2 and two $AdS_3$ (Schwarzschild and RN) black holes in section 3. Aspects of a quintessence underlying a $D$-instanton are qualitatively discussed in a brane/anti-brane pair production. We conclude by summarizing the results in section 4. 

\section{Quantum geometries on ${\mathbf{AdS_4}}$ brane}
A geometric torsion underlying a $U(1)$ gauge theory on a $D_4$-brane may alternately be viewed vacuum created pair of $(D{\bar D})_3$-brane. The non-linear charge pumped by a dynamical two form is believed to incorporate non-trivial effective geometries on a brane and an anti-brane in a second order curvature formalism \cite{spsk-JHEP}. In fact a pair production underlying a brane and anti-brane universe(s) began at a cosmological horizon with a Big Bang singularity. A conservation of linear momentum at the pair production primarily ensures that a brane universe is moving away from an anti-brane universe in opposite direction and vice-versa. Their separation distance grows with time since its inception at a Big Bang. It implies that 
an effective torsion curvature formalism on a $D_4$-brane underlie the notion of a vacuum created pair of $(D{\bar D})_3$-brane which are believed to be separated by a large extra dimension in the present day cosmology. 

\sp
\noindent
In other words an observer in a $4D$ brane universe is unaware of the fifth dimension. The hidden dimension to a brane or anti-brane universe has been shown to possess its origin in a non-perturbative torsion curvature on a $D_4$-brane. A repulsive gravity sourced by an axion underlying a quintessence has recently been argued to drift away a brane universe from its anti-brane \cite{psskk}. The growth in transverse length scale may justify the observed acceleration of the universe in  cosmic microwave background radiations (CMB). The hot de Sitter brane/anti-brane vacua give away radiations and cool down upon their expansion. Presumably the stringy modes underlying a torsion \cite{candelas-HS} decouple via a series of geometric transitions from the hot brane/anti-brane vacua and the reduced geometries are believed to approach Einstein vacua in the formalism. It may imply that the expansion of a brane universe is in conundrum with a low energy limit defined by a large $r$ in a brane or an anti-brane world-volume.

\subsection{$\mathbf{(D{\bar D})}$-instanton pair}
An axion in the Ramond-Ramond sector of type IIA or IIB superstring theory describes a D-instanton. Importantly a D-instanton is known to incorporate a non-perturbative correction in a perturabtive string theory. In fact a pair creation of brane/anti-brane universe has been argued to begin instantaneously with a Big Bang at a cosmological horizon presumably underlying an effective de Sitter background on a $D_4$-brane. It provokes thought to believe that a pair of brane/anti-brane universe in the present day cosmology had began with a $(D{\bar D})$-instanton at an early epoch in de Sitter underlying a geometric torsion formalism \cite{spsk-JHEP}. The expansion of a pair of brane and anti-brane universe from its origin at a Big Bang was argued to go through a series followed by the nucleation of higher dimensional pairs such as a $(D{\bar D})$-string and a $(D{\bar D})$-membrane. The nucleation lead to an expansion of the brane universe. The process naturally slows down with the emission of radiations and it may have stopped with a vacuum created $(D{\bar D})_3$-brane  with a large extra dimension transverse to both the world-volumes. Needless to mention that the prescription completely fills a $D_4$-brane world-volume with a pair of $(D{\bar D})_3$-brane. The hidden fifth dimension is believed to break the supersymmetry 
inherent to a BPS $D_4$-brane and may be described by a non-BPS brane configuration.

\sp
\noindent
In other words, an instanton local degree on an anti $D_3$-brane influences a vacuum created $D_3$-brane within its pair. Arguably the instantaneous affect of a ${\bar D}$-instanton on a $D_3$-brane is in agreement with the non-perturbative universe in $4D$. In the context an effective five dimensional torsion curvature has been constructed in a second order formalism by the authors in the recent past \cite{spsk-JHEP,spsk-PRD,spsk-NPB-P,spsk-jaat,spsk-NPB1}. In particular we have explored a two form $U(1)$ gauge dynamics on a $D_5$-brane in presence of a background open string effective metric. It  was shown that a global NS two form underlying the open string metric can source a pure de Sitter vacuum. A vacuum pair of $(D{\bar D})_3$-brane has been argued to be created, at the cosmological horizon in a background de Sitter, by a two form quanta in a non-linear $U(1)$ gauge theory on a $D_5$-brane. The pair production mechanism is inspired by the novel idea of Hawking radiation at the event horizon of a semi-classical black hole by a photon in a $U(1)$ gauge theory \cite{hawking}. An underlying electric non-linear $U(1)$ charge in the effective curvature formalism incorporates an extended particle/anti-particle pair creation in the quantum theory. The assertion for a pair of (non-fundamental) string may be confirmed with a local two form on a $D_5$-brane. The stringy nature further reassures an underlying quantum gravity phenomenon in Yang-Mills theory \cite{thooft,kar-maharana,hanada}. Generically a higher form Poincare dual to an one form may be considered on an appropriate $D_p$-brane for $p>3$ underlying a background de Sitter vacuum to describe higher dimensional vacuum pairs such as a membrane/anti-membrane, a three brane/anti-brane and other higher dimensional brane pairs. It may be recalled that a space filling brane ($D_9$-brane) in type IIB superstring theory is indeed described by the closed string (quantum gravity) modes there. A vacuum created pair $(D{\bar D})_3$ is described with an extra fifth transverse dimension between a brane and an anti-brane and may be viewed as a space filling $D_5$-brane. More over the $U(1)$ gauge invariance of the two form on a $D_5$-brane has been exploited to enforce the metric fluctuations in the five dimensional effective curvature formalism \cite{spsk-JHEP}. It was shown that the brane universe created, at the cosmological horizon with a Big Bang, on a pair of $(D{\bar D})_3$-brane may be viewed through a low energy perturbative string vacuum in presence of a quantum correction underlying a (non-perturbative) $D$-brane or an anti $D$-brane. 

\sp
\noindent
Generically a vacuum created pair of $(D{\bar D})_3$-brane requires a vacuum geometry which should be protected by a horizon of a black hole. We shall note that under a gauge choice for a global NS two form, an AdS background (open string effective) metric may be obtained in a non-linear gauge theory on a $D_5$-brane. Formally the AdS background may be viewed as a pure de Sitter across the cosmological horizon under a flip of light-cone \cite{spsk-JHEP}. It may also be understood via an electric-magentic duality on a $D_3$-brane. In fact the conserved mass may be identified with the square of electric and magnetic non-linear charges respectively for de Sitter and AdS. Thus the emergent quantum geometries created on a pair of $(D{\bar D})_3$-brane may be alternately be viewed through a pair of a negative mass brane and a positive mass anti-brane or vice-versa.

\sp
\noindent
Alternately, the quantum fluctuations were shown to be sourced by a geometric torsion ${\cal H}_3$ in an effective curvature scalar ${\tilde{\cal K}}$ formalism. A gauge theoretic torsion is modified to a geometric torsion, $i.e.\ H_3\rightarrow {\cal H}_3$, in presence of a new connection which is believed to modify the covariant derivative. The new connection underlying an effective curvature has been identified with a geometric torsion. A detailed investigation leading to a geometric torsion has been discussed by the authors in \cite{spsk-JHEP}. A geometric torsion in the formalism may be given by
\bea
{\cal H}_{\mu\nu\lambda}&=&3{\cal D}_{[\mu}B_{\nu\lambda ]}\nonumber\\
&=&3\nabla_{[\mu}B_{\nu\lambda ]} + 3{{\cal H}_{[\mu\nu}}^{\alpha}
{B^{\beta (NS)}}_{\lambda ]}\ g_{\alpha\beta}
\nonumber\\ 
&=&H_{\mu\nu\lambda} + \left ( H_{\mu\nu\alpha}{B^{(NS)\alpha}}_{\lambda} + \rm{cyclic\; in\;} \mu,\nu,\lambda \right )\ +\ H_{\mu\nu\beta} {B^{(NS)\beta}}_{\alpha} {B^{(NS)\alpha}}_{\lambda} + \dots \;\ \label{geometricT-1} \\
{\rm where}\quad {\cal D}_{\lambda}B_{\mu\nu}&=&\nabla_{\lambda}B_{\mu\nu} + {1\over2}{{{\cal H}}_{\lambda\mu}}^{\rho}B^{(NS)}_{\rho\nu} 
- {1\over2}{{\cal H}_{\lambda\nu}}^{\rho}B^{(NS)}_{\rho\mu}\ .\label{geometricT-2}
\eea
A commutator, of covariant derivatives, was worked out to explore the possibility of an effective curvature in a second order formalism for a Mankowski metric $g_{\mu\nu}$. It is given by
\be
\Big [ D_{\mu}\ ,\ D_{\nu} \Big ]A_{\lambda}=\ {{\tilde{\cal K}}_{\mu\nu\lambda}}^{\rho}A_{\rho} \ +\ {\cal F}_{\nu\mu}A_{\lambda}\ +\ 
{{\cal H}_{\mu\nu}}^{\rho}\ {\cal D}_{\rho}A_{\lambda}\ .\label{geometricT-3}
\ee 
\bea
{\rm Where}&&{4{\tilde{\cal K}}_{\mu\nu\lambda}}^{\rho}=2\partial_{\mu}{{\cal H}_{\nu\lambda}}^{\rho} -2\partial_{\nu} {{\cal H}_{\mu\lambda}}^{\rho} 
+ {{\cal H}_{\mu\lambda}}^{\sigma}{{\cal H}_{\nu\sigma}}^{\rho}-{{\cal H}_{\nu\lambda}}^{\sigma}{{\cal H}_{\mu\sigma}}^{\rho}\nonumber\\
{\rm and}&&{\cal F}_{\mu\nu}=\ F_{\mu\nu}\ +\ {\cal H}_{\mu\nu}^{\lambda}{\cal A}_{\lambda}\ .\label{gauge-12}
\eea
The fourth order space-time curvature tensor is anti-symmetric under an interchange of a pair of indices, $i.e.\ {\tilde{\cal K}}_{\mu\nu\lambda\rho}
=-{\tilde{\cal K}}_{\nu\mu\lambda\rho}=-{\tilde{\cal K}}_{\mu\nu\rho\lambda}$. The effective curvature tensor incorporates the dynamics a geometric torsion in five dimensions. Interestingly for a non propagating torsion, in a gauge choice, the effective curvature tensor reduces to the Riemannian tensor: ${\tilde{\cal K}}_{\mu\nu\lambda\rho}\rightarrow R_{\mu\nu\lambda\rho}$. The gauge choice ensures a decoupling of a dynamical two form in a perturbative gauge theory which in turn dissociates a non-perturbative ($D$-brane) correction from the low energy string vacuum. The gauge fields in a $D$-brane world-volume theory satisfy:
\bea
&&{\cal D}_{\lambda}{\cal F}^{\lambda}_{\mu\nu}=0\nonumber\\
{\rm and} &&\partial_{\lambda}H^{\lambda\mu\nu}\ +\ \Big ( g^{\alpha\beta}\partial_{\lambda}\ g_{\alpha\beta}\Big )
H^{\lambda\mu\nu}=0\ . \label{gauge-12a}
\eea
Now we consider a vacuum created pair of $(D{\bar D})_3$-brane universe at a (cosmological) horizon in a non-linear gauge theory on a $D_5$-brane. The presence of horizon ensures that the created vacuum pair can not undergo annihilation. An extra fifth dimension, hidden to the created pair, is reassured by considering ${\tilde{\cal K}}$ on $S^1$. Then the effective dynamics may seen to be described by two independent $4D$ geometric curvatures: ${\cal K}$ and ${\cal F}_2$. The absence of a scalar dynamics in $4D$ may be checked from the fact that 
${\tilde{\cal K}}$ possesses its origin in a two form gauge theory. A count on local degrees on a $D_5$-brane and that on a vacuum pair of $(D{\bar D})_3$-brane further reassures the absence of a dilaton dynamics. For $\kappa^2=(2\pi)^{5/2}g_s\alpha'$, the effective curvatures on a pair of $(D{\bar D})_3$-brane may be given by
\be
S= {1\over{3\kappa^2}}\int d^4x {\sqrt{-\det G_{\mu\nu}}}\;\;\ \left ( {\cal K}\ -\ {3\over4} {\bar{\cal F}}_{\mu\nu}
{\cal F}^{\mu\nu} \right )\ .\label{gauge-11}
\ee
In absence of a geometric torsion, the geometric two form in eq(\ref{gauge-12}) reassures a gauge theoretic non-linear field strength on a $D_3$-brane. Intuitively, a geometric torsion incorporates a spin angular momentum into the vacuum created pair. It implies that a spinning motion on a vacuum created $D_3$-brane is in opposite direction to that on anti $D_3$-brane. In a low energy limit the spin slows down significantly and may be ignored to retain the BPS configuration, $i.e.$ a BPS brane and an anti BPS brane. In the limit the extra fifth dimension becomes large and hence a brane universe and an anti-brane universe may appear to be independent of each other. In the case the quintessence field dynamics freezes on an anti brane. Then the brane universe is precisely governed by two local degrees of a gauge theoretic non-linear one form.

\subsection{Gauge field ansatz}
The two form ansatz on a vacuum created $D_3$-brane is given by
\bea
B^{(NS)}_{t\theta}&=& B^{(NS)}_{r\theta} = {{b}\over{\sqrt{2\pi\alpha'}}}\ ,\nonumber\\
B_{\theta\phi}&=& {{p}\over{\sqrt{2\pi\alpha'}}} \sin^2\t = \left ({{p}\over{b}}\right ) B_{r\t}\ \sin^2\t\ , \nonumber\\
{\cal A}_t&=& {{Q_e}\over{r}}\qquad\; {\rm and}\qquad {\cal A}_{\phi}\ =\ {{-Q_m}\over{\sqrt{2\pi\alpha'}}} \cos\t\ ,\label{ansatz-1}
\eea
where $(b,p)>0$ are constants and $(Q_e,Q_m)$ denote the non-linear (electric, magnetic) charges. The gauge choice re-assures a vanishing torsion: $(H_3=0={\cal H}_3)$. A geometric two form field strength (\ref{gauge-12}) in the gauge choice reduces to a
non-linear gauge theoretic $F_2$ and its nontrivial components are given by
\bea
F_{0r} &=& {{Q_e}\over{r^2}}\nonumber \\
{\rm and}\qquad F_{\t\phi}&=& {{Q_m}\over{2\pi\alpha'}}\sin\t\ .\label{ansatz1a}
\eea
They satisfy the field equations of motion (\ref{gauge-12a}). The local degrees in ${\cal A}_{\mu}$ ensures that a propagating geometric torsion in a five dimensional effective curvature scalar ${\tilde{\cal K}}$ formalism. However a constant torsion on a vacuum created $D_3$-brane or an anti $D_3$-brane may imply that the emergent quantum geometry is free form torsion and may well be described by a four dimensional Einstein vacuum. A nontrivial torsion is known to incorporate a quantum (metric) fluctuation. A priori, a torsion-less brane universe may formally be identified with a low energy string vacuum in the formalism. However analysis reveals that the open string effective metric sourced by a global mode of NS two form may seen to receive a geometric correction due to the non-linear $U(1)$ charges underlying a generic ${\cal F}_2$ on a $D_3$-brane. Generically the emergent metric may be expressed as:
\be
{G}_{\mu\nu}=\left ( g_{\mu\nu}\ +\ B^{(NS)}_{\mu\lambda}g^{\lambda\rho}B^{(NS}_{\rho\nu}\ \pm\ {\bar{\cal F}}_{\mu\lambda}g^{\lambda\rho}{\bar {\cal F}}_{\rho\nu}\right )\ .\label{emetric}
\ee
A priori the effective geometries obtained on a vacuum created brane for $(2\pi\alpha')=1$ are given by
\bea
ds^2=&-&\left (1 + {{b^2}\over{r^2}} \pm {{Q_e^2}\over{r^4}}\right ) dt^2 + \left ( 1 - {{b^2}\over{r^2}} \pm {{Q_e^2}\over{r^4}}\right ) dr^2 - {{2b^2}\over{r^2}} dt dr \nonumber \\ &&-\ {{2bp\sin^2\theta}\over{r^2}}\left ( dt + dr \right )d\phi+\left ( 1 - {{p^2\sin^2 \theta \pm Q_m ^2 }\over{r^4}}\right )  r^2 d\Omega^2\ .\label{ads-1}
\eea 
It may be noted that a dynamical ${\cal A}_{\mu}$, on a vacuum created $D_3$-brane, possesses its origin in a five dimensional torsion underlying an effective curvature scalar ${\tilde{\cal K}}$. The metric fluctuations sourced by ${\cal F}_2$ in eq(\ref{emetric}) further ensure a non-perrturbative ($D$-brane) correction and hence the solutions (\ref{ads-1}) may be describe a quantum vacuum.
\subsection{Schwarzshild black hole}
Now we perform an appropriate Weyl scaling of the emergent metric to access an AdS window on a vacuum created pair of $(D{\bar D})_3$-brane.
The scaling may be given by
\be
G_{\mu\nu}= {{-b^2}\over{r^2}} {\tilde G}_{\mu\nu}\ .\qquad\qquad {}\label{ads-2}
\ee
Then the scaled effective geometries in a brane window, $i.e.\ N_e<r<b$ with $N_e^4<<r^4<<b^4$, may be approximated to yield:
\bea
ds^2 &=&\left (1+ {{r^2}\over{b^2}} \pm {{N_e^2}\over{r^2}}\right ) dt^2 +  \left (1 + {{r^2}\over{b^2}} \pm {{N_e^2}\over{r^2}}\right )^{-1} dr^2 + 2dt dr \nonumber \\ && +\ {{2p}\over{b}}\left ( dt + dr \right )\sin^2\t\ d\phi +\ {1\over{b^2}}\left (p^2 \sin^2\theta \pm Q_m^2- r^4\right )d\Omega^2\ ,\label{ads-3}
\eea
where $N_e=(Q_e/b)$ signifies a modified electric charge. The scaling ensures: $0<b<\infty$. The line-element implies that a magnetic non-linear charge decouples from the $S_2$ symmetric line-element and hence $Q_m$ becomes redundant to the emergent geometries on a brane. In a global scenario, $i.e.$ on a vacuum created pair of $(D{\bar D})_3$-brane, some of the brane charges are nullified by the anti-brane. The quantum geometry is essentially sourced by a two form on a $D_4$-brane and is given by
\bea
ds^2&=&\left(1+ {{r^2}\over{b^2}} \pm {{N_e^2}\over{r^2}}\right) dt^2 +  \left(1 + {{r^2}\over{b^2}} \pm {{N_e^2}\over{r^2}}\right)^{-1} dr^2 + {{2p}\over{b}} \sin^2\t\ d\phi dt\nonumber \\
 &&\qquad\qquad\qquad\qquad\qquad\qquad\qquad\qquad\qquad +\ {1\over{b^2}}\left(p^2 \sin^2\theta - r^4\right)d\Omega^2\ .\label{ads-4}
\eea 
It signifies the vital role played by a background NS two form to define a rotating AdS brane underlying a vacuum created pair of $(D{\bar D})_3$-brane. Analysis reveals an additional upper bound on the radial coordinate, $i.e.\; r<\sqrt{p\sin\theta}$. It ensures that the upper bound on $r$ takes a maximum value at the equatorial plane and the brane geometry does not access to the poles. Thus for $p\neq 0$, the effective quantum geometries describe an AdS Schwarzschild $4D$ quantum black hole on a pair of $(D{\bar D})_3$-brane. The remaining brane geometry may not be preferred to describe our universe appropriately due to the cosmic censorship hypothesis. For a real angular momentum the effective AdS Schwarszchild quantum black hole on a $D_3$-brane within a pair is given by
\bea
ds^2&=&-\left (1+ {{r^2}\over{b^2}} - {{N_e^2}\over{r^2}}\right ) dt^2\ + \ \left (1 + {{r^2}\over{b^2}} - {{N_e^2}\over{r^2}}\right )^{-1} dr^2 +\ {{2p}\over{b}} \sin^2\t\ d\phi dt \ +\ \rho^2 d\Omega^2\ ,\label{ads-5}
\eea
where $\rho= {1\over{b}}{\sqrt{p^2 \sin^2\theta - r^4}}$. Unlike to an AdS Schwarzschild in Einstein vacuum the emergent AdS brane quantum black hole may a priori seen to be associated with a spin angular momentum. Nevertheless the spin, intrinsic to the emergent geometry, is sourced by a background NS two form presumably on an anti $D_3$-brane. A non-propagating torsion in a gauge choice (\ref{ansatz-1}) ensures a Riemannian curvature 
(${{\cal K}_{\mu\nu\lambda}}^{\rho}\rightarrow {R_{\mu\nu\lambda}}^{\rho}$) in the formalism. Naively the curvature scalar $R$ may seen to possess a singularity in a limit $r\rightarrow 0$. However the curvature singularity is not accessible due to a lower bound on $r$ in a brane universe.

\sp
\noindent
In a low energy limit the fifth extra dimension between a brane and anti-brane becomes large. The background spin may be ignored in the limit to describe a typical AdS Schwarzschild on an effective  $D_3$-brane. In the case the non-linear $N_e^2$ may define the mass of an AdS black hole and $b$ denotes an AdS radius. The effective gravitational potential hints at a hidden fifth large dimension in the low energy vacuum. Thus an AdS Schwarzschild quantum black hole may be realized through an appropriate brane window.

\subsection{Black hole bound state} 
In this section we Weyl scale the emergent geometry on a brane to address a new AdS black hole underlying a brane and an anti-brane in the formalism.
In the case the scaling may be identified as:
\be
G_{\mu\nu}= {{Q_e^2}\over{r^4}} {\tilde G}_{\mu\nu}\ .\qquad\qquad {}\label{ads-6}
\ee
The scaled effective geometries in a brane window, $i.e.\ r<N_e$ with $r^8<<N_e^4$, may be approximated to yield:
\bea
ds^2&=&-\left (1+ {{r^2}\over{N_e^2}} + {{r^4}\over{Q_e^2}}\right ) dt^2 + \left (1 + {{r^2}\over{N_e^2}} - {{r^4}\over{Q_e^2}}\right )^{-1} dr^2 -\ {{2b^2r^2}\over{Q_e^2}}\ drdt\nonumber \\  &&-\ {{2bp}\over{Q_e^2}} r^2 (dt + dr) \sin^2\t\ d\phi\ +\ {1\over{Q_e^2}}\left (r^4-p^2 \sin^2\theta - Q_m^2\right ) r^2 d\Omega^2\ .\label{ads-7}
\eea
Unlike to an effective AdS Schwarzschild black hole in eq(\ref{ads-5}), the magnetic non-linear charge $Q_m$ does not decouple from the $S_2$-symmetric line-element. A magnetic charge along with a background $B_2$-fluctuation tend to decrease the effective radius of $S_2$. The magnitude of the redefined electric charge $|N_e|$ defines the AdS radius in a brane universe. In a global scenario the $4D$ vacuum reduce to yield:
\bea
ds^2&=&-\left (1+ {{r^2}\over{Q_e^2}}\left (b^2 + r^2\right ) \right )dt^2 -\ {{2bp}\over{Q_e^2}}\ r^2 \sin^2\t\ d\phi dt + 
\left (1 + {{r^2}\over{Q_e^2}}\left (b^2 - r^2\right )\right )^{-1} dr^2 \nonumber \\ && + \ {1\over{Q_e^2}}\left (r^4-p^2 \sin^2\theta - Q_m^2\right ) r^2 d\Omega^2\ .\label{ads-8}
\eea
The brane geometry may be re-expressed with the re-defined electric and magnetic non-linear charges. It is given by
\bea
ds^2&=&-\left (1+ {{r^2}\over{N_e^2}} + {{r^4}\over{Q_e^2}}\right ) dt^2 + \left (1 + {{r^2}\over{N_e^2}} - {{r^4}\over{Q_e^2}}\right )^{-1} dr^2 -\ {{2}\over{N_e^2}}\left ( {{p}\over{b}}\right )\ r^2 \sin^2\t\ d\phi dt\nonumber\\ && +\ {1\over{N_e^2}}\left ({{r^4}\over{b^2}}- {{p^2}\over{b^2}} \sin^2\theta - N_m^2\right ) r^2 d\Omega^2\ .\label{ads-8a}
\eea
The new emergent quantum geometry is characterized by an event horizon at $r\rightarrow r_e$ underlying an effective AdS quantum black hole. It is computed to yield:
\be
r_e={{b}\over{\sqrt{2}}}\left (1 + \sqrt{1 +{{4Q_e^2}\over{b^4}}} \right )^{1/2}\ .\label{ads-10}
\ee
For a large $b$, defined with $b^4>>Q_e^2$, the event horizon may be approximated by $r_b\approx b$. Similarly for a large $Q_e$, defined with $Q_e^2>>b^4$, the event horizon approximates to $r_Q={\sqrt{Q_e}}$. In addition, the AdS brane is described by a spin angular momentum which is primarily sourced by a background NS two form fluctuation. The torsion free brane universe allows one to compute the Ricci scalar $R$. Analysis reveals a curvature singularity in a limit $r\rightarrow 0$ in a brane window. 

\sp
\noindent
Interestingly the emergent AdS quantum black hole on a brane may be re-expressed as an $AdS$ with non-perturbative quantum corrections. Naively it is given by
\bea
ds^2&=&-\left (1+ {{r^2}\over{N_e^2}}\right ) dt^2 +   \left (1 + {{r^2}\over{N_e^2}}\right )^{-1} dr^2 - {{2bp}\over{Q_e^2}} r^2 \sin^2\t\ d\phi dt \nonumber \\ &&+\ \left (r^4 - p^2 \sin^2\theta - Q_m^2\right ) {{r^2}\over{Q_e^2}} d\Omega^2  + \ {{r^4}\over{Q_e^2}}\left (-dt^2 + dr^2\right ) \ .\label{ads-10a}
\eea
The quantum correction is associated with a flat metric and hence is free from any space-time curvature in the formalism. The non-perturbative quantum term has been shown to possess its origin in an extra dimension between a brane and anti-brane \cite{spsk-NPB1, spsk-NPB2}. The correction term may appropriately be identified for $Q_m\neq 0$ to yield:
\bea
ds^2&=&-\left (1+ {{r^2}\over{N_e^2}}\right ) dt^2 +  \left (1 + {{r^2}\over{N_e^2}}\right )^{-1} dr^2-\ {{2bp}\over{Q_e^2}} r^2 \sin^2\t\ d\phi dt \nonumber \\ && +\ \left (1+{{p^2}\over{Q_m^2}} \sin^2\theta \right )r^2 d\Omega^2+\ {{r^4}\over{Q_e^2}}\left (-dt^2\ +\ dr^2\ +\ r^2 d\Omega^2 \right )\ .\label{ads-10b}
\eea
In absence of a background fluctuation, $i.e.\ p=0$, a non-linear magnetic charge plays an important role by incorporating an $S_2$-symmetric line element into the $AdS_4$ brane. It takes a form:
\bea
ds^2&=&-\left (1+ {{r^2}\over{N_e^2}}\right ) dt^2\ +\ \left (1 + {{r^2}\over{N_e^2}}\right )^{-1} dr^2\ + \  r^2 d\Omega^2 +\ {{r^4}\over{Q_e^2}}\ ds^2_{\rm flat}\ .\label{ads-10c}
\eea
It implies that an $AdS$ quantum black hole may alternately be viewed through a non-perturbative quantum correction to an effective $AdS_4$-brane. 
A global mode of a NS two form is sourced by a parameter $b$. It does not seem to play a significant role in the brane geometry except for its background spin. For $b\rightarrow 0$ and $p\rightarrow 0$, the brane world black hole (\ref{ads-8}) in a global scenario may be simplified to yield:
\bea
ds^2&=&-\ \left (1+ {{r^4}\over{Q_e^2}}\right )dt^2\ +\  \left (1 - {{r^4}\over{Q_e^2}}\right )^{-1} dr^2 \ +\ \left (1 - {{r^4}\over{Q_m^2}}\right ) r^2 d\Omega^2\ .\label{ads-10d}
\eea 
A priori, the Ricci scalar is computed for the brane geometry to confirm a curvature singularity at $r\rightarrow 0$. An additional curvature singularity  may be seen to appear at $r\rightarrow {\sqrt{Q_m}}$ for a non self-dual electric and magnetic field. Nevertheless the brane universe is protected 
by an event horizon at $r\rightarrow r_Q={\sqrt{Q_e}}$ for $|Q_e|>|Q_m|$. In other words the new emergent geometry describes a quantum black hole which may be viewed as a trapped regime. The fact that the new quantum geometry is not asymptotically flat imply that the brane universe may not describe by an isolated system. The upper bound on $r$ due to the non-linear electric charge in the formalism underlying a quantum black hole on a brane may not possess an analogue solution in Einstein gravity. The new quantum geometry in a brane universe may describe a phase in the early epoch. Interestingly the Ricci scalar is computed to confirm a positive constant vacuum energy density at the horizon $r\rightarrow {\sqrt{Q_e}}$. It may describe a quantum de Sitter 
with a cosmological horizon at $r\rightarrow {\sqrt{Q_e}}$ on a pair of $(D{\bar D})_3$-brane. The gauge choice in the case ensures a flat background metric in the gauge theory on a  $D_4$-brane.

\sp
\noindent
On the other hand the brane universe (\ref{ads-8}) may be analyzed for $p=0$. It is given by 
\bea
ds^2&=&-\left (1+ {{r^2}\over{N_e^2}} + {{r^4}\over{Q_e^2}}\right ) dt^2 +  \left (1 + {{r^2}\over{N_e^2}} - {{r^4}\over{Q_e^2}}\right )^{-1} dr^2 +\ \left (1 - {{r^4}\over{Q_m^2}}\right ) r^2 d\Omega^2\ .\label{ads-11}
\eea
It may describe a new quantum vacuum leading to an AdS black hole on a vacuum created pair of $(D{\bar D})_3$-brane. The curvature singularity at $r\rightarrow 0$ in the brane universe is protected by an event horizon $r_e$ obtained in eq(\ref{ads-10}). 

\sp
\noindent
The quantum geometries on a $D_3$-brane,  underlying an AdS Schwarzschild (\ref{ads-5}) and an AdS new black hole (\ref{ads-11}), may be analyzed with a single scale along its radial coordinate. The bound $N_e<r_I<b$ in a brane window-I and $r_{II}<N_e$ in a brane window-II respectively define an AdS Schwarzschild and an AdS new black hole for a large $b$. They define two different regimes on a $D_3$-brane. The bounds imply that an AdS new quantum black hole (\ref{ads-11}) is realized on a pair of $(D{\bar D})_3$-brane for relatively small $r$ than that for an AdS Schwarzschild black hole (\ref{ads-5}). It may imply that an AdS new black hole may be viewed as a non-perturbative bound state which separates itself from a continuous AdS Schwarzschild black hole spectrum. Notionally the emerging scenario is in conformity with the idea of a mass gap between a bound state $AdS_3$ and a continuous BTZ black hole \cite{BTZ}.

\subsection{Quintessence in torsion curvature}
Formally one may redefine the radial coordinate $r^2= R>0$ for large length scale underlying an expanding universe. Then the effective $AdS_3$ brane (\ref{ads-11}) may be re-expressed for as:
\bea
ds^2&=&-4R\left (1+ {{R}\over{N_e^2}} + {{R^2}\over{Q_e^2}}\right ) dt^2 +\ 4\left (1 -{{R^2}\over{Q_m^2}}\right ) R^2 d\Omega^2 +  \left (1 + {{R}\over{N_e^2}} - {{R^2}\over{Q_e^2}}\right )^{-1} dR^2 \ .\label{ads-12}
\eea
The linear term $R$ in the causal patches may argued to be sourced by a quintessence scalar \cite{huan-jun-quintessence}. In the brane/anti-brane scenario a quintessence is believed to be described by an axion which in turn is Poincare dual to a torsion. In the context an axion sources a $D$-instanton which may be a potential candidate to describe the dark energy in the brane universe. The quadratic term in $R$ would like to be sourced by a cosmological constant in the brane world formalism. A varying curvature scalar ensures a non constant vacuum energy density. It signifies the role of a quintessence in the brane universe.

\sp
\noindent
Interestingly a brane universe (\ref{ads-12}) may seen to be sourced by a quintessence. In a brane window $0<R<N_e^2$, the geometry may be approximated to yield:
\bea
ds^2&=&ds_I^2 \ +\ ds_{II}^2\ ,\nonumber\\
{\rm where}\quad ds_I^2&=&-4R\left (1+ {{R}\over{N_e^2}}\right ) dt^2+ \left (1 + {{R}\over{N_e^2}}\right )^{-1} dR^2 + 4R^2 d\Omega^2\nonumber\\
{\rm and}\quad ds_{II}^2&=& - {{\Lambda}\over{3}}R^2 \left ( -4R dt^2 + dR^2+ 4R^2 d\Omega^2\right )\ ,\label{ads-13}
\eea
where $\Lambda$ denotes the cosmological constant. It shows that an effective AdS brane universe underlying $ds_I^2$ is primarily sourced by a quintessence underlying a non-perturbative curvature in the formalism. The remaining line element $ds_{II}^2$ apparently describes a geometric correction with a coupling formally defined with $\Lambda$. The scalar curvatures associated with the line elements $ds_I^2$ and $ds_{II}^2$ are computed to yield: $${\cal R}_{I}=\ {{-3}\over{R^2}}\left ( 1 + {{3R}\over{N_e^2}}\right )\qquad {\rm and}\qquad  {\cal R}_{II}= {{-3}\over{R^2}}\ .$$ 
Varying curvature scalars ensure the role of quintessence in the formalism. They appear to blow up in a limit $R\rightarrow 0$. However a brane universe is defined with $R>0$ and hence it does not access the curvature singularity. The negative scalar curvature further reassures an effective AdS brane. Thus an electric non-linear charge is primarily responsible for a non-perturbative $D_3$-brane world. It ensures a varying vacuum energy density underlying the dark energy in brane universe.

\sp
\noindent
In the context a time variation of cosmological vacuum energy density in the early epoch leading to an effective de Sitter geometry has been discussed in the formalism \cite{psskk}. A quintessence is known to describe a time varying energy density. In the recent years the observational cosmology primarily focuses on the measurement of $\omega$ sourced by a quintessence. It takes a bounded value: $-1<\omega<-1/3$. The inaccessible axion to an observer in an effective space-time on a $D_3$-brane provokes thought to identify the Poincare dual of torsion as a quintessence axion in the formalism. The idea of a quintessence energy density sourced by an axion field on an anti-brane is remarkable. A quintessence axion is believed to be a potential candidate to describe the dark energy in our universe. The fact that the dark energy permeats all the space and tends to enhance the rate of expansion of universe may be understood via an extra transverse dimension between a pair of vacuum created $D_3$-brane and ${\bar D}_3$-brane in the formalism. 

\section{Quantum patches on ${\mathbf{AdS_3}}$ brane}
We explore a plausible scenario underlying a brane universe in $3D$ in a torsion curvature formalism. An effective $AdS_2$ may seen to emerge on a vacuum created pair of $(D{\bar D})_2$-brane primarily sourced by a two form in a $U(1)$ gauge theory on a $D_4$-brane. Alternately we consider an effective torsion curvature scalar ${\tilde{\cal K}}$ underlying a $D_4$-brane on $S^1\times S^1$. The notion of a pair of lower dimensional brane/anti-brane vacuum created by a non-linear gauge field on a higher dimensional $D_p$-brane shall be explored in the section. We investigate the plausible dynamical aspect of an effective gravity in a lowest possible space-time dimension on a $D_2$-brane.

\subsection{Dynamical axion}
An axion in a vacuum created anti $D_3$-brane may seen to be sourced by a torsion using Poincare duality. Thus a non-linear axionic charge is vital to the emergent geometries on a vacuum created $D_3$-brane within a pair. Very recently an axionic scalar in the formalism is conjectured to 
describe a quintessence which is known to incorporate a varying vacuum energy density \cite{psskk}. Thus the dark energy in a brane universe may be sourced by a dynamical axion on an anti-brane.

\sp
\noindent
Now we focus on a vacuum created pair of $(D{\bar D})_2$-brane from a $D_3$-brane or an anti $D_3$-brane followed from a $D_4$-brane.
A gauge field ansatz on a vacuum created pair of $(D{\bar D})_2$-brane is worked underlying the field equations of motion (\ref{gauge-12a}). The gauge fields are given by
\bea
&&B^{(NS)}_{0r}=\left ({{r^2}\over{b^2}}-{{2M}\over{r}}\right)^{1/2}\ ,\nonumber\\
&&B_{r\phi}=\left ({{2r^4}\over{b^2}}-4Mr\right)^{1/2}= r{\sqrt{2}}\ B^{(NS)}_{0r}\nonumber\\
{\rm and}&& A_0^i\ =\ - Q^i \ln r\ ,\label{3d-1}
\eea
where $i=(1,2,3)$ signify three independent dynamical one forms or axion fields on a $D_2$-brane. The background NS two form components are described by two independent parameters $(M,b)>0$ with $r^3>(2Mb^2)$ on a brane. The global modes of NS two form in an open string is known to describe an effective gravity on an anti $D_2$-brane which in turn describes the background to a $D_2$-brane universe. The gauge choice ensures a torsionless $H_3=0$ effective geometry in presence of the non-linear axionic charges $Q^i$ on a $D_2$-brane. An axion in a RR-sector sources a $D$-instanton. A vacuum created pair of $(D{\bar D})$-instanton is argued to incorporate a non-perturbative curvature instantaneously on a brane/anti-brane universe. Thus the role of an instanton is believed to serve as a fundamental building block to the emergent space-time in a brane \cite{spsk-JHEP,psskk}.

\sp
\noindent
The nontrivial $U(1)$ field strength $F^i_{0r}= Q^i/r$ may be checked to satisfy the equation of motion: $\nabla_{\lambda}F^{\lambda\mu}_i=0$. Apparently the effective geometry on a vacuum created $D_2$-brane with a pair of $(D{\bar D})_2$-brane may be obtained using the emergent metric (\ref{emetric}). The line element is given by
\bea
ds^2&=&-\left( 1+ \frac{r^2}{b^2} - \frac{2M}{r} \pm \frac{Q_i^2}{r^2} \right) dt^2  + \left(1- \frac{r^2}{b^2} +\frac{2M}{r} \pm \frac{Q_i^2}{r^2}  \right)dr^2  \nonumber \\  &&+\ 2{\sqrt{2}} \left({{r^2}\over{b^2}}-\frac{2M}{r}\right ) r d\phi dt  + \left ( 1 -{{2r^2}\over{b^2}} + {{4M}\over{r}}\right )r^2 d\phi^2 \ .\label{3d-2}
\eea
In a brane window for $(M, |Q_i|)<r<b$, the effective geometries may be approximated with ($M^2<<r^2$, $Q_i^4<<r^4<<b^4$) to yield:
\bea
ds^2&=&-\left( 1+ \frac{r^2}{b^2} - \frac{2M}{r} \pm \frac{Q_i^2}{r^2} \right) dt^2 + \left(1+ \frac{r^2}{b^2} -\frac{2M}{r} \mp \frac{Q_i^2}{r^2}  \right)^{-1}dr^2 \nonumber \\  &&+\ 2{\sqrt{2}} \left({{r^2}\over{b^2}}-\frac{2M}{r}\right ) r d\phi\ dt +\ \left ( 1 -{{2r^2}\over{b^2}} + {{4M}\over{r}}\right )r^2 d\phi^2\ .\label{3d-3}
\eea
Primarily the dynamical aspects ($Q_i\neq 0$) is incorporated by an one form gauge field on a vacuum created pair of $(D{\bar D})_2$-brane. The remaining two parameters ($M,b$) describe the background charges on a three dimensional brane universe. Interestingly the background geometry to a $D_2$-brane is assumed to be an effective AdS Schwarzschild on an anti $D_2$-brane within a vacuum created pair. The background AdS Schwarzschild is obtained by a  geometric engineering dictated via an ansatz (\ref{3d-1}) in the formalism. The background spin is characterized by an angular velocity:  $\Omega^{\phi}_{BG}=-{\sqrt{2}}/(3r_H)$ where $r_H$ denotes the horizon radius in the AdS Schwarzschild black hole on an anti $D_2$-brane. 

\subsection{Discrete transformation of causal patches}
The absence of a local torsion in the gauge choice further ensures a Riemannian curvature and hence formally: ${\cal K}^{(3)}\rightarrow R^{(3)}$. For instance, three local degrees in a generic curvature scalar ${\tilde{\cal K}}$ on $S^1\times S^1$ may be manifested in the $F^i_2$ on a $D_2$-brane with a non-propagating torison leading to a Riemannian curvature on an anti $D_2$-brane in the formalism. Since a nonzero vacuum energy density plays a significant role to describe a $3D$ Einstein gravity \cite{BTZ}, it may be pumped into an anti $D_2$-brane through a dynamical axion on a $D_2$-brane
within a pair. As a result the hidden extra dimension within a pair of $(D{\bar D})_2$-brane plays a vital role. The extra dimension is further 
re-assured by an effective gravitational potential in the background AdS Schwarzschild black hole. Intuitively one may consider a typical AdS Schwarzschild black hole in $4D$ Einstein gravity with a cosmological constant which may serve as a background metric to a $D_2$-brane. Analysis reveals that the non-linear gauge dynamics on a $D_2$-brane underlie a background Einstein gravity. It is in conformity with the generic notion of a $D$-brane dynamics established in type IIA or IIB  superstring theory. Presumably the interpretation validates the notion of a vacuum created pair of lower dimensional brane/anti-brane by a higher form.

\sp
\noindent
The mixed causal patches in the quantum geometries may be separated out using a ($2\times 2$) projection matrix ${\cal M}$ as discussed in refs \cite{spsk-JHEP,spsk-PRD,spsk-NPB1,spsk-NPB2,psskk}. The matrix takes a form:
\begin{equation}
{\cal M}=\frac{1}{2}\left( \begin{array}{ccc}
{{G}_{tt}^+} && {{G}_{rr}^+}\\ && \\
{{G}_{rr}^-} && {{G}_{tt}^-}
\end{array} \right)\ .\label{3d-4}
\end{equation}
Explicitly, the causal components $G^{\pm}_{\mu\nu}$ are defined as:
\bea
G_{tt}^{\pm}&=& -\left( 1+ \frac{r^2}{b^2} - \frac{2M}{r} \pm \frac{Q_i^2}{r^2} \right) \nonumber \\
{\rm and}\quad G_{rr}^{\pm}&=& \left(1+ \frac{r^2}{b^2} -\frac{2M}{r} \pm \frac{Q_i^2}{r^2}  \right)^{-1}\ .\label{3d-5}
\eea
The inverse matrix becomes
\begin{equation}
{\cal M}^{-1}=\frac{1}{2\det{\cal M}}\left( \begin{array}{ccc}
{{G}_{tt}^-} && {-{G}_{rr}^+}\\ && \\
{-{G}_{rr}^-} && {{G}_{tt}^+}
\end{array} \right)\ ,\label{3d-6}
\end{equation}
It may be checked that the matrix (\ref{3d-4}) is projected on the column vectors:
$$\pmatrix{{1}\cr {0}}\qquad {\rm and}\qquad \pmatrix{{0}\cr {1}}$$
to yield the brane geometries obtained in eqs(\ref{3d-3}). The matrix determinant is computed in the regime to yield:
\be
{\det {\cal M}}= \left ({{r^2}\over{b^2}}- {{2M}\over{r}}\right )\ .
\ee
The determinant is evaluated at the event horizon $r_h$ of a background black hole to re-assure $\det {\cal M}_h=-1$. It implies that a discrete transformation using an inverse matrix (\ref{3d-6}) projection on the column vectors incorporates correct sign in the causal quantum patches. It leads to two distinct AdS black holes on a vacuum created pair of $(D{\bar D})_2$-brane. Considering a real angular momentum, 
the effective AdS brane geometries in $3D$ are given by
\bea
ds^2_{\rm Sch}&=&-\left( 1+ \frac{r^2}{b^2} - \frac{2M}{r} -\frac{Q_i^2}{r^2} \right) dt^2  +\ \left(1+ \frac{r^2}{b^2} -\frac{2M}{r} - \frac{Q_i^2}{r^2}  \right)^{-1}dr^2 \nonumber \\ &&+\ 2{\sqrt{2}} \left({{r^2}\over{b^2}}-\frac{2M}{r}\right ) r d\phi\ dt+ \left ( 1 -{{2r^2}\over{b^2}} + {{4M}\over{r}}\right )r^2 d\phi^2\ .\label{3d-7}
\eea
\bea
{\rm and}\qquad ds^2_{\rm RN}&=&-\left( 1 + \frac{r^2}{b^2} - \frac{2M}{r}\ + \frac{Q_i^2}{r^2} \right) dt^2+ \left(1+ \frac{r^2}{b^2} -\frac{2M}{r} + \frac{Q_i^2}{r^2}  \right)^{-1}dr^2 \nonumber \\ && +\ 2{\sqrt{2}} \left({{r^2}\over{b^2}}-\frac{2M}{r}\right ) r d\phi\ dt + \left ( 1 -{{2r^2}\over{b^2}} + {{4M}\over{r}}\right )r^2 d\phi^2\ .\label{3d-8}
\eea
\subsection{Characteristic features}
The effective AdS brane geometries may seen to be characterized by a spin angular momentum. They are computed at their event horizon(s) and may be expressed as:
\bea
\Omega^{\phi}_{\rm Sch}= {{-{\sqrt{2}}\left (r_e^2-Q_i^2\right )}\over{r_e\left (3r_e^2-2Q_i^2\right )}}\qquad {\rm and} \qquad \Omega^{\phi}_{\rm RN}= {{-{\sqrt{2}}\left (r_+^2+Q_i^2\right )}\over{r_+\left (3r_+^2+2Q_i^2\right )}}\ .\label{spin}
\eea
The Ricci curvature scalar, underlying the torsion free quantum geometries, blows up at $r\rightarrow 0$. However a lower bound on $r$ in a brane universe does access the curvature singularity otherwise present in Einstein vacuum. A lower bound on r further reassures an ultraviolet cutoff and hence a quantum black hole in the formalism. The emergent geometries (\ref{3d-7}) and (\ref{3d-8}) are 
characterized by one and two horizons respectively. They describe an effective AdS Schwarzschild black hole (\ref{3d-7}) and an 
AdS Reissner-Nordstrom black hole (\ref{3d-8}) on a vacuum created $D_2$-brane or an anti-brane.

\sp
\noindent
On the other hand, a single horizon implies that the effective gravitational potential sourced by a background NS two form ($b\neq 0$) and a dynamical one form ($Q_i\neq 0$) are not independent of each other. The background AdS black hole mass $M$ may be redefined to absorb $Q_i$. The renormalized ''mass'' is computed at the black hole horizon $r_H$ on a brane to yield:
\bea
M_n&=&M \left (1 + {{Q_i^2}\over{2Mr_H}}\right )\nonumber\\
&=&r_H\left (1 + {{r_H^2}\over{b^2}} + {{Q_i^2}\over{r_H^2}}\right )\ .\label{massBH}
\eea 
Then the black hole in eq(\ref{3d-7}) corresponds to an effective AdS Schwarzschild on a vacuum created $D_2$-brane within a pair. The brane universe may be approximated to yield:
\bea
ds^2&=&-\left( 1+ \frac{r^2}{b^2} - \frac{2M_n}{r}\right) dt^2\ - \ 2r{\sqrt{2}}\ d\phi dt +\ \left(1+ \frac{r^2}{b^2} -\frac{2M_n}{r}\right)^{-1}dr^2 \ +\ 3 r^2 d\phi^2\ .\label{3d-9}
\eea
The effective gravitational potential in the AdS Schwarzschild black hole hints at a fourth dimension transverse to a vacuum created $D_2$-brane. Alternately the Schwarzschild geometry may be re-expressed with the renormalized charges ${\cal Q}^i$. The background black hole mass $M$ may be absorbed to describe the renormalized charge. It may be given by
\be
{\cal Q}_i=Q_i\left (1 + {{2Mr_H}\over{Q^2_i}}\right )^{1/2}\ .\label{chargeBH}
\ee 
The black hole may be (\ref{3d-7}) may be re-expressed to describe an effective gravity in a gauge theory on a $D_2$-brane. It is given by
\bea
ds^2&=&-\left( 1+ \frac{r^2}{b^2} - \frac{{\cal Q}_i^2}{r^2}\right) dt^2\ +\ \left(1+ \frac{r^2}{b^2} -\frac{{\cal Q}_i^2}{r^2}\right)^{-1}dr^2 - 2r{\sqrt{2}}\ d\phi dt\ +\ 3 r^2 d\phi^2\ .\label{3d-10}
\eea
The effective AdS Schwarzschild black hole in $3D$ hints at an underlying gauge theory on a $D_4$-brane. In a global scenario a charge due to a brane annihilates that on its anti-brane to yield:
\bea
ds^2&=&-\left( 1+ \frac{r^2}{b^2} - \frac{{\cal Q}_i^2}{r^2}\right) dt^2\ +\ \left(1+ \frac{r^2}{b^2} -\frac{{\cal Q}_i^2}{r^2}\right)^{-1}dr^2+\ 3 r^2 d\phi^2\ .\label{3d-11}
\eea
A priori the parameter $M$, in a background NS two form ansatz, does not seem to play a role in a global scenario underlying a vacuum created pair of 
$(D{\bar D})_2$-brane. Analysis reveals that the parameter $M$ plays a vital role to define a background AdS black hole which in turn is used by a non-linear gauge field to produce a pair of brane/anti-brane. 

\section{Concluding remarks}
A torsion curvature formalism, underlying a global NS two form coupled to a local two form in a $U(1)$ gauge theory, on a $D_4$-brane was investigated for some of the effective AdS geometries. Quintessence energy density was argued to be described by a $D$-instanton in the formalism which in turn was shown  to be sourced by a torsion.

\sp
\noindent
In particular the quantum geometries were described on a vacuum created pair of $(D{\bar D})_3$-brane at an event horizon. 
We have obtained an $AdS_4$ Schwarzschild black hole and an $AdS_4$ black hole bound state respectively in two distinct windows on a $D_3$-brane within a pair. An axion, underlying a $D$-instanton, was primarily argued to describe a quintessence in the formalism. In particular an axion is intrinsic to a torsion on an anti $D_3$-brane which in turn influences a $D_3$-brane universe via an extra dimension. The curvature scalars were computed for the effective AdS black holes to confirm the varying vacuum energy density in a brane universe. 

\sp
\noindent
In addition we have obtained quantum black holes leading to an $AdS_3$ Schwarschild and an $AdS_3$ Reissner-Nordstrom on a vacuum created pair of $(D{\bar D})_2$-brane. A $D_2$-brane within a pair was shown to be described by three non-linear axionic charges. The significance of an axion on a $D_2$-brane or on an anti-brane was analyzed to enhance our understanding on dark energy underlying an accelerating brane universe. 


\def\anp{Ann. of Phys.}
\def\cmp{Comm.Math.Phys.\ {}}
\def\prl{Phys.Rev.Lett.\ {}}
\def\prd#1{{Phys.Rev.}{\bf D#1}}
\def\jhep{JHEP\ {}}{}
\def\cqg#1{{Class. \& Quant.Grav.}}
\def\plb#1{{Phys. Lett.} {\bf B#1}}
\def\npb#1{{Nucl. Phys.} {\bf B#1}}
\def\mpl#1{{Mod. Phys. Lett} {\bf A#1}}
\def\ijmpa#1{{Int.J.Mod.Phys.}{\bf A#1}}
\def\mpla#1{{Mod.Phys.Lett.}{\bf A#1}}
\def\rmp#1{{Rev. Mod. Phys.} {\bf 68#1}}
\def\jaat{J.Astrophys.Aerosp.Technol.\ {}} {}


\end{document}